\documentclass{article}
\usepackage{ijcai19}

\usepackage{times}
\usepackage{soul}
\usepackage{url}
\usepackage[hidelinks]{hyperref}
\usepackage[utf8]{inputenc}
\usepackage[small]{caption}
\usepackage{graphicx}
\usepackage{amsmath}
\usepackage{booktabs}
\usepackage{mathtools,amssymb,amsthm,amsfonts,amsmath}
\usepackage{pifont}
\usepackage{bbm}
\usepackage{tcolorbox}
\usepackage{tikz,tikz-qtree}
\usetikzlibrary{arrows,automata,shapes,positioning,calc} 
\usepackage{enumitem}
\usepackage{stmaryrd}
\usepackage{forest}
\usepackage{wrapfig}
\usepackage{graphicx}
\usepackage[mathscr]{euscript}
\usepackage{tipa}
\usepackage{gb4e}

\renewcommand{\succ}{\ensuremath\lhd}

\colorlet{good}{blue!75}
\colorlet{bad}{red!75}

\newcommand{\tuple}[1]{\ensuremath{\langle #1 \rangle}}

\newcommand{\defeq}{\overset{def}{=} }

\newtheorem{theorem}{Theorem}
\newtheorem{definition}{Definition}

\title{Tensor Product Representations of Subregular Formal Languages}

\author{
    Jonathan Rawski
    \affiliations
    Department of Linguistics and Institute for Advanced Computational Science \\ Stony Brook University, USA \emails
    jonathan.rawski@stonybrook.edu
}

\begin{document}

\maketitle

\begin{abstract}
  This paper provides a geometric characterization of subclasses of the regular languages. We use finite model theory to characterize objects like strings and trees as relational structures. Logical statements meeting certain criteria over these models define subregular classes of languages. The semantics of such statements can be compiled into tensor structures, using multilinear maps as function application for evaluation. This method is applied to consider two properly subregular languages over different string models.
\end{abstract}

\section{Introduction}
Formal language theory provides a way to explicitly tie the complexity of linguistic patterns to specific claims about memory organization and thus provides an indirect way of measuring the cognitive demands of language. The Chomsky hierarchy \cite{chomsky59} is the standard way in formal language theory to measure language complexity. 
One particular class of interest is the regular languages. One of the most well-studied objects in computer science, the regular languages are characterized by regular expressions, finite-state machines, and statements in Monadic Second-Order logic, among others \cite{Thomas1997}. Linguistically, the regular class has been shown to sufficiently characterize phonological and morphological patterns \cite{kaplan-kay1994}. 

More recently, it has been shown that many linguistic patterns inhabit subclasses of the regular languages (see \cite{Heinz-2018-CNPG} for an overview). These subclasses each have particular automata-theoretic, algebraic, and logical characterizations \cite{Rogers-HeinzEtAl-2013-CSC}. This paper considers logical characterizations of subregular language classes over various representations of structures that are defined using finite model theory.  

In the last few decades, finite model theory has emerged as a powerful description language for linguistics. Model theory has been used for comparisons of particular grammatical theories in phonology and syntax \cite{rogers1998descriptive,Pullum2007,graf2010thesis}, and for studying the nature of linguistic structures and processes themselves \cite{Heinz-2018-CNPG}. In these formulations, linguistic structures like strings and trees are modeled using relational information which holds among the items characterizing a particular domain. 

Constraints over model-theoretic structures, as well as translation between one structure and another, can be described using statements in mathematical logic. Such constraints and transformations express in an elegant way the relationship between grammars and representation. In particular, various types of logic over specific model-theoretic representations (say, strings and trees) yield particular classes of grammars \cite{rogers1996strict}. 

It is of interest to see how these models may be characterized geometrically. Geometric approaches to language and symbolic cognition in general have become increasingly popular during the last two decades. There is work dealing with conceptual spaces for sensory representations \cite{gardenfors2004conceptual}, multilinear representations for compositional semantics \cite{blutner2009concepts,aerts2009quantum}, and dynamical systems for modeling language processes \cite{beim2008language,tabor2009dynamical}. 



One particularly significant contribution in this area is Smolensky’s Tensor Product Representations \cite{smolensky1990tensor}. Here, subsymbolic dynamics of neural activation patterns in a vector space description become interpreted as symbolic cognitive computations at a higher-level description by means of ``filler/role" bindings via tensor products. These tensor product representations form the symbolic foundation of Harmonic Grammar and Optimality Theory, and have been successfully employed for phonological and syntactic computations. \cite{Smolensky-Legendre-HMV-1}. 

In tensor product representations, symbolic structures are decomposed into structural roles and fillers, bound together using the tensor product. For example, strings can be decomposed into a tensor realizing string positions, each of which is bound to a tensor realizing different symbols in some alphabet. Similarly, tree structures can be represented recursively, with a tensor product representation using tree node position as the structural role, and using an entire subtree as an alphabet filler symbol. \cite{hale2001parser} use this tree representation to describe a Harmonic Grammar for context-free languages, and \cite{beim2012geometric} use them to formalize \cite{Stabler1997}'s Minimalist Grammars. 

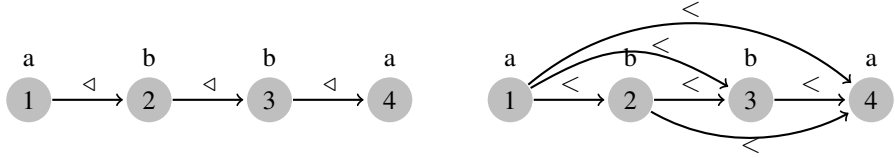
\begin{figure*}
    \centering
    \begin{tikzpicture}[shorten >=1pt,->,thick, scale=0.8]
      \tikzstyle{vertex}=[circle,fill=black!25,minimum size=17pt,inner sep=0pt]
      \draw node[vertex] [label={[align=center]above:a}] (1) at (0,0) {1};
      \draw node[vertex][label={[align=center]above:b}] (2) at (2,0) {2};
      \draw node[vertex][label={[align=center]above:b}] (3) at (4,0) {3};
      \draw node[vertex][label={[align=center]above:a}] (4) at (6,0) {4};
      \draw (1) -- (2); \node at (1,0.25) {$\triangleleft$};
      \draw (2) -- (3); \node at (3,0.25) {$\triangleleft$};
      \draw (3) -- (4); \node at (5,0.25) {$\triangleleft$};
      
      \draw node[vertex][label={[align=center]above:a}] (s) at (8,0) {1};
      \draw node[vertex][label={[align=center]above:b}] (r) at (10,0) {2};
      \draw node[vertex][label={[align=center]above:b}] (i) at (12,0) {3};
      \draw node[vertex][label={[align=center]above:a}] (S) at (14,0) {4};
      \draw (s) -- (r); \node at (9,0.25) {$<$};
      \draw (r) -- (i); \node at (11,0.25) {$<$};
      \draw (i) -- (S); \node at (13,0.25) {$<$};
      \draw (s) .. controls +(+30:2cm) and +(+150:2cm) .. (i);
      \draw (s) .. controls +(+40:2.5cm) and +(+140:2.5cm) .. (S);
      \draw (r) .. controls +(-30:2cm) and +(-150:1cm) .. (S);
      \node at (10.5,.9)  {$<$};
      \node at (11,1.5)   {$<$};
      \node at (12,-0.75) {$<$};
    \end{tikzpicture}
    \caption{Visualizations of the successor (left) and precedence (right) models of $abba$.}
    \label{fig:cm}
    \vspace{-2em}
\end{figure*}

While tensor product representations form a powerful method for geometrically interpreting symbolic structures, explicitly modeling the relational structures given by model-theory, as well as the logical constraints characteristic of subregular languages, is an open issue. This paper provides this connection, by translating model-theoretic structures into vector spaces and describing logical grammatical constraints over them using tensors.    

There has been some recent work on embedding logical calculi using tensors. \cite{grefenstette2013towards} introduces tensor-based predicate calculus that realizes logical operations. \cite{yang2014embedding} introduce a method of mining Horn clauses from relational facts represented in a vector space. \cite{serafini2016learning} introduce logic tensor networks that integrate logical deductive reasoning and data-driven relational learning. \cite{sato2017embedding} formalizes Tarskian semantics of first-order logic in vector spaces. Here we apply Sato's method for translating model-theoretic representations and first-order logic into tensors. 

Finally, tensor methods and subregular grammars/automata have been used to evaluate and interpret neural networks (see \cite{rabanser2017introduction}). \cite{Avcu+2017-SCDL} tested the generalization capacity of LSTM networks on the Strictly Local and Strictly Piecewise languages. \cite{mccoy2018rnns} showed that recurrent neural networks (RNNs) implicitly encode tensor product representations, and \cite{weiss2017extracting} used regular languages to test the generalization capacity of RNNs. Explicit translation of subregular languages into tensors over various representations thus allows model-theoretic linguistics to study neural nets in a principled way.    


\section{Model Theoretic Representations}
\subsection{Elements of Finite Model Theory}


Model theory, combined with logic, provides a powerful way to study and understood mathematical objects with structures~\cite{Enderton2001}.  
This paper only considers finite relational models~\cite{Libkin2004}.
\begin{definition}
A \emph{model signature} is a tuple $S=\tuple{D;R_1,R_2,\ldots,R_m}$ where the domain $D$ is a finite set, and each $R_i$ is a $n_i$-ary relation over the domain. 
\end{definition}
In this paper, the relations are at most binary. 

\begin{definition}
A \emph{model for a set of objects} $\Omega$ is a total, one-to-one function from $\Omega$ to structures whose type is given by a model signature.
\end{definition}

\subsection{String Models}
We can view strings as models. The set of all possible finite strings of symbols from a finite
alphabet $\Sigma$ and the set of strings of length $\leq n$ are
$\Sigma^*$ and $\Sigma^{\leq n}$, respectively. The unique empty
string is represented with $\lambda$. The length of a string $w$ is
$|w|$, so $|\lambda| =$ 0. 
If $u$ and $v$ are two strings then we denote their concatenation
with $uv$. If $w$ is a string and $\sigma$ is the $i$th symbol in $w$ then $w_i=\sigma$, so $abcd_2=b$.

A conventional model for strings in $\Sigma^*$ is given by the signature $\Gamma^\succ=\tuple{D;\succ, [R_\sigma]_{\sigma\in\Sigma} }$ and the function $M^\succ: \Sigma^*\to \Gamma^\succ$ such that $M^\succ(w) = \tuple{D^w;\succ, [R^w_\sigma]_{\sigma\in\Sigma} }$
 where $D^w= \{1,\ldots, |w|\}$ is the domain, $\succ=\{(i,i+1)\in D\times D \mid j=i+1\}$ is the successor relation which orders the elements of the domain, and $[R^w_\sigma]_{\sigma\in\Sigma}$ is a set of $|\Sigma|$ unary relations such that for each $\sigma\in\Sigma$, $R^w_\sigma=\{ i\in D^w\mid w_i=\sigma\}$.
We will usually omit the superscript $w$ since it will be clear from the context.

For example, with $\Sigma=\{a,b,c\}$ and the model above for strings, we have 
\begin{multline}
M^\succ(abba)= \big\langle D=\{1,2,3,4\};\\ \succ=\{(1,2),(2,3),(3,4)\};\\ R_a=\{1,4\}; R_b=\{2,3\}, R_c=\emptyset\big\rangle
\label{eqn:succabba}
\end{multline}
Figure~\ref{fig:cm} illustrates $M^\succ(abba)$ on the left.
 
Another conventional model is the precedence model, with the signature $\Gamma^< =\tuple{D;<, [R_\sigma]_{\sigma\in\Sigma} }$. 
It differs from the successor model only in that the order relation is defined with general precedence ($<$), which is defined as $< =\{(i,j)\in D\times D\mid i < j\}$  \cite{Buchi1960,McNaughtonPapert1971,Rogers-HeinzEtAl-2013-CSC}. 
Under this model, the string $abba$ has the following model.

\begin{multline}
M^<(abba) = \big\langle D=\{1,2,3,4\};\\ 
<=\{(1,2),(1,3), (1,4), (2,3), (2,4), (3,4)\},\\
R_a=\{1,4\}, R_b=\{2,3\}, R_c=\emptyset\big\rangle
\end{multline}

Figure~\ref{fig:cm} illustrates $M^{<}(abba)$ on the right.

Under both model signatures, each string $w \in \Sigma^*$ of length $k$ has a unique interpretable structure.
The model of string $w = \sigma_1\sigma_2\ldots\sigma_k$ has domain $D = \{1; 2\ldots k\}$, and for each $\sigma \in \Sigma, R_\sigma =
\{i \in D | w_i = \sigma\}$. The difference between $M^\succ$ and $M^\prec$ is the ordering relation. Under the successor
model $M^\succ$, the ordering relation is $\succ \defeq \{(i,i+1) \in D x D\}$, while for the precedence model $M^\prec$, the
ordering relation is $\prec \defeq \{(i,j) \in D \times D | i \prec j\}$

However, structures are more general in that they correspond to any mathematical structure conforming to the model signature. 
As such, while a model of a string $w$ will always be a structure, a structure will not always be a model of a string $w$. 
The \emph{size} of a structure $S$, denoted $|S|$, coincides with the cardinality of its domain.

\section{Logic, Languages, and Language Classes}
\label{sec:grammars}

Usually a model signature provides the vocabulary for some logical language $\mathscr{L}$, which contains $N$ constants
$\{e_1,\ldots, e_N\}$. 
Following notation of \cite{sato2017embedding}, a model $M = (D,I)$ is thus a pair of domain, a nonempty set $D$ and an interpretation 
$I$ that maps constants $e_i$ to elements
(entities, individuals) $I(e_i) \in D$ and $k$-ary predicate symbols
$r$ to $k$-ary relations $I(r) \subseteq D^k$

An assignment $a$ is a mapping from variables $x$ to an element $a(x) \in D$. It
provides a way of evaluating formulas containing free variables. Syntactically terms mean variables and/or constants and atomic formulas or atoms $r(t_1,\ldots,t_k)$ are comprised of a $k$-ary predicate symbol $r$ and $k$ terms $t_1,\ldots,t_k$ some of
which may be variables. Formulas $F$ in $\mathscr{L}$ are inductively
constructed as usual from atoms using logical connectives
(negation $\neg$, conjunction $\land$, disjunction $\lor$) and quantifiers
$(\exists,\forall)$.

There are several well-known connections between logical statements and languages classes. 
Most famous is \cite{Buchi1960}'s result that languages characterizable by finite-state machines, the regular languages, are equivalent to statements in Monadic Second-Order Logic over the precedence model for strings (and successor, since precedence is MSO-definable from successor). 

\begin{figure}[htbp!]
    \centering
    \includegraphics[width=\linewidth]{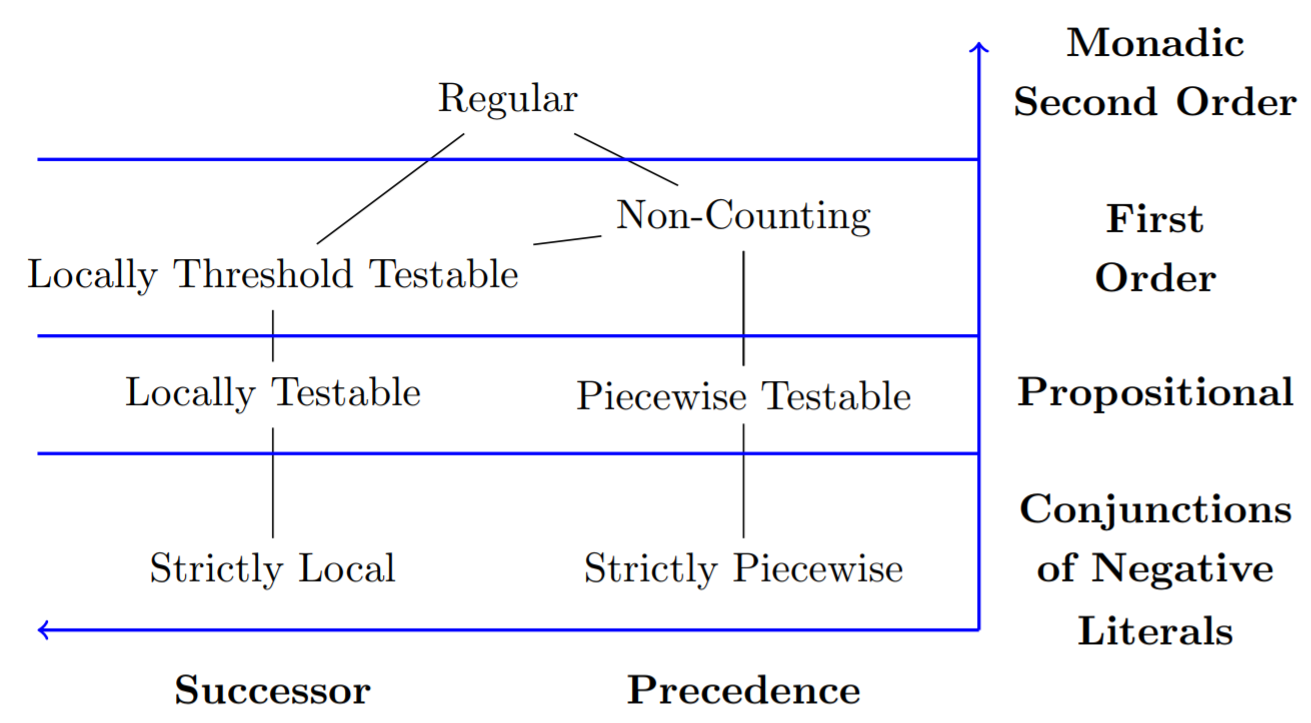}
    \caption{Subregular Hierarchy of Languages}
    \label{fig:SRLogic}
\end{figure}

Within the regular languages, many well-known subregular classes can be characterized by weakening the logic \cite{McNaughtonPapert1971,Rogers-HeinzEtAl-2013-CSC,Thomas1997}. An overview of these connections is shown is Figure~\ref{fig:SRLogic}  

Here we restrict ourselves to first-order logic, as it is the lightest restriction corresponding to properly subregular languages. First-order statements over particular model signatures define distinct language classes. For the successor model, 
\cite{Thomas1997} characterizes FO(\succ) in terms of Local Threshold Testability, equivalence in terms of the multiplicity of $k$-subfactors up to some fixed finite threshold $t$.

\begin{theorem}{(\cite{Thomas1997})}
A set of strings is First-order definable over$\tuple{D;\succ, [R_\sigma]_{\sigma\in\Sigma}}$  iff it is Locally Threshold Testable.
\end{theorem}

Correspondingly, First-order formulas over the precedence model characterize the Non-Counting or Star-Free class of languages. 

\begin{theorem}{(\cite{McNaughtonPapert1971})}
Languages that are first-order definable over $\tuple{D;\prec, [R_\sigma]_{\sigma\in\Sigma}}$ are Non-Counting.
\end{theorem}

Non-Counting languages are those languages definable in linear temporal logic, and which have with aperiodic syntactic monoids. Since the Successor relation $(\succ)$ is first-order definable from Precedence $(\prec)$, the Non-Counting class properly includes the Locally Threshold Testable class.

Further restrictions of the logic over these signatures to propositional logic or conjunctions of negative literals characterize subclasses of the LTT and Star Free languages, yielding the Local and Piecewise hierarchies. \cite{Rogers-HeinzEtAl-2013-CSC}.

\section{Tensor Representations of First-Order Logic}
This section overview's \cite{sato2017embedding}'s method for embed a model domain and signature into a vector space, using tensors to encode relational information.

 Scalars are denoted with lower 
case letters like $a$. Vectors mean column vectors and we denote 
them by boldface lower case letters like \textbf{a} and 
\textbf{a}'s components by $a_i$. $\mathscr{D}^{\prime}= 
\{\mathbf{e}_1,\ldots,\mathbf{e}_N\}$ is the
standard basis of $N$ -dimensional Euclidean space $\mathbb{R}^{N}$
where $\mathbf{e}_{i} = (0 \cdots,1,\cdots,0)^{T}$ is a vector that
has one at the $i$ -th position and zeros elsewhere. Such vectors 
are called one-hot vectors. \textbf{1} is a vector of all ones. 
We assume square matrices, written by boldface upper 
case letters like \textbf{A}. \textbf{I} is an identity matrix, and
$\mathbbm{1}$ is a matrix of all ones. Order-p tensors $\mathcal{A} \in \mathbb{R}^{D^p}$, are also denoted by $\left\{a_{i_{1},\ldots,i_{p}\}}\right\} (1\leq$ $i_{1},\ldots,i_{p}\leq N) .$ $\mathcal{A}$'s component $a_{i_{1},\ldots,i_{p}}$ is also written as $( \mathcal{A})_{i_{1},\ldots,i_{p}} .$ 
 $( \mathbf{a} \bullet \mathbf{b} ) = \mathbf{a}^{T} \mathbf{b}$ is the inner product of \textbf{a} and $\mathbf{b}$ whereas $\mathbf{a} \circ \mathbf{b} = \mathbf{ab}^{T}$ is their outer product. $\mathbf{1}\circ\cdots\circ\mathbf{1}$ is a $k$-order tensor, and $\mathbf{1} \circ \cdots \circ \mathbf{1} \left( \mathbf{e}_{i_{1}} , \ldots , \mathbf{e}_{i_{k}} \right)=(\mathbf{1} \bullet \mathbf{e}_{i_{1}}) \cdots( \mathbf{1}\bullet \mathbf{e}_{i_{k}}) = 1$.

There exists an isomorphism between tensors and multilinear maps \cite{bourbaki1989commutative}, such
that any curried multilinear map
\begin{equation}\nonumber
    f : V _ { 1 } \rightarrow \ldots \rightarrow V _ { j } \rightarrow V _ { k }
\end{equation}
can be represented as a tensor $\mathcal{T}_f \in V_k \otimes V_j \otimes \ldots \otimes V_1$. This means that
tensor contraction acts as function application.
This isomorphism guarantees that there exists such a
tensor $\mathcal{T}^f$ for every $f$, such for any $v_1 \in V_1,\ldots, v_j \in V_j$:
\begin{equation}
    f \mathbf { v } _ { 1 } \ldots \mathbf { v } _ { j } = \mathbf { v } _ { k } = \mathcal { T } ^ { f } \times \mathbf { v } _ { 1 } \times \ldots \times \mathbf { v } _ { j }
\end{equation}

 Following \cite{sato2017embedding}, we first isomorphically map a model $M$ to a model $M^\prime$ in $\mathbb{R}^N$. We map entities $e_i \in D$ to one-hot vectors \textbf{e_i}. So $D$ is mapped to $D^\prime = \{\mathbf{e}_1,\ldots,\mathbf{e}_N\}$, the basis of $\mathbb{R}^N$. We next map a $k$-ary relation $r$ in $M$ to a $k$-ary relation $r^\prime$ over $D^\prime$ which is computed by an order-$k$ tensor $\mathcal{R} = \{r_{i_1,\ldots,i_k}\}$, whose truth value $\llbracket r(e_{i_1},\ldots,e_{i_k})\rrbracket$ in $M$ is given by

$\llbracket r(e_{i_1},\ldots,e_{i_k})\rrbracket $
\vspace{-1em}
\begin{equation}
\begin{aligned}
&= \mathcal{R}( \mathbf{e}_{i_{1}},\ldots,\mathbf{e}_{i_{k}}) \\
&= \mathcal{R}\times_{11} \mathbf{e}_{i_{1}}\times_{12} \cdots \times_{1,i_{k}} \mathbf{e}_{i_{k}}\\
&= r_{i_{1},\ldots ,i_{k}} \in\{1,0 \}\left(\forall i_{1},\ldots,i_{k} \in\{1,\ldots,N\} \right)
\end{aligned}
\end{equation}

We identity $r^\prime$ with $\mathcal{R}$ so that $\mathcal{R}$ encodes the $M$-relation $r$. Let $M^\prime$ be a model $(D^\prime ,I^\prime )$ in $\mathbb{R}^N$ such that $I^\prime$ interprets entities by $I^\prime(e_i) = e_i (1 \leq i \leq N)$ and relations $r$ by $I^\prime(r) = \mathcal{R}$. 

For the purposes of this paper, we restrict ourselves to binary relations and predicates.  When r is a binary predicate, the corresponding tensor $\mathcal{R}$ is a bilinear map and represented by an adjacency matrix \textbf{R} as follows:
\begin{equation}
    \llbracket(e_{i},e_{j}) \rrbracket = (\mathbf{e}_{i} \cdot \mathbf{Re}_{j})=\mathbf{e}_{i}^{T}\mathbf{R} \mathbf{e}_{j}=r_{ij} \in \{1,0\}
\end{equation}
Note that when $r(x,y)$ is encoded by $\mathcal{R}$ as $(\mathbf{x\bullet R y})$, $r(y,x)$ is encoded by
$\mathbf{R}^T$, since $(\mathbf{y\bullet R x}) = (\mathbf{x}\bullet \mathbf{R}^T \mathbf{y})$ holds

We next inductively define the evaluation $\llbracket F \rrbracket_{I^\prime,a^\prime}$ of a formula $F$ in $M$. Let $a$ be an assignment in $M$ and $a^\prime$ the corresponding assignment in $M^\prime$, so $a(x) = e_i$ iff $a^\prime (x)= \mathbf{e}_i$. For a ground atom $r(e_{i_1},...,e_{i_k})$, define
\begin{multline}
    \llbracket r(e_{i_{1}}, \ldots,e_{i_{k}}\rrbracket ^ \prime  = \underline {\mathbf{R}} (\mathbf{e}_{i_{1}}, \ldots,\mathbf{e}_{i_{k}}) \\ (\forall i_{1},\ldots,i_{k} \in \{1,\ldots,N \}).
\end{multline}
where $\mathcal{R} = \{r_{{i_1},...,{i_1}}\}$ is a tensor encoding the $M$-relation $r$ in $M$. By definition $\llbracket F\rrbracket_{I,a} = \llbracket F\rrbracket_{I,a}$
holds for any atom $F$. Negative literals are evaluated using $\neg \mathcal{R}$ defined as

\begin{equation}
    \begin{aligned} \llbracket \neg r \left(e_{i_{1}}, \ldots,e_{i_{k}} \rrbracket ^ { \prime } \right. &= \neg \mathcal{R} \left( \mathbf{e}_{i_{1}} , \ldots , \mathbf{e}_{i_{k}} \right) \\ 
    \text{where}~\neg \mathcal{R} &\defeq  \overbrace {\mathbf{1} \circ \cdots \circ \mathbf{1}}^{k}-\mathcal{R}
    \end{aligned}
\end{equation}

$\neg \mathcal{R}$ encodes an $M$-relation $\neg r_1$. Negation other than negative literals, conjunction, disjunction, and quantifiers are evaluated in $M^\prime$ as follows.

\begin{align}
\llbracket \neg F \rrbracket^{\prime} &= 1 - \llbracket F \rrbracket^{\prime} \\
\llbracket F_{1} \land \cdots \land F_{h} \rrbracket^{\prime} &= \llbracket F_{1} \rrbracket^{\prime}\cdots \llbracket F_{h} \rrbracket^{\prime} \\
\llbracket F_{1}\lor\cdots\lor F_{h}\rrbracket^{\prime} &= \min_1 (\llbracket F_{1}\rrbracket^\prime +\ldots+\llbracket F_{h} \rrbracket^\prime)  \\  
\llbracket \exists y F \rrbracket^{\prime} &= \min_1(\sum_{i=1}^{N} \llbracket F_{y\leftarrow e_{i}} \rrbracket^{\prime})
\end{align}

Here the operation $\min_1 (x) = \min(x,1) = x$ if $x < 1$, otherwise 1, as componentwise application. $F_{y\leftarrow e_i}$
means replacing every free occurrence of $y$ in $F$ with $e_i$. We treat universal quantification is treated as
$\forall x F = \neg \exists x \neg F$.

\section{Compiling Subregular Formulas into Tensor Algebra}

\cite{sato2017embedding} presents an algorithm for compiling any first-order formula into a tensor embedding without grounding. The algorithm works by converting a formula into prenex normal form, Each quantified statement is put in conjunctive or disjunctive normal form, depending on the quantifier, and each formula is then converted to the appropriate tensor realization. 

Here we restrict ourselves to formulas with binary predicates, which as stated above may be represented as adacency matrices since their corresponding tensor is a bilinear map. Sato shows that in that in these  cases, we can often “optimize” compilation by directly compiling a formula $F$ using matrices.

\subsection{Compiling a Locally Threshold Testable Formula}
Here we demonstrate a formula which is properly First Order over the successor model for strings, characterizing a Locally Threshold Testable Language. With the ability to distinguish distinct occurrences of a symbol we can define a formula which is satisfied by strings containing exactly one occurrence of some symbol $b$. Such a system is seen in phonological stress patterns in the world's languages, which often mandate exactly one primary stress in a word. We do this by asserting that there is some position in which $b$ occurs $\exists x (R_b(x))$, and that there are no other positions in which $b$ occurs $\land(\forall y)[R_b(y) \rightarrow (x = y)]]$. The conjunction of these two gives the FO formula
\begin{equation}
    F_{\text{one-}B} = (\exists x \forall y)[R_b (x) \land [R_b(y) \rightarrow (x = y)]]
\end{equation}

Converting this into prenex normal form we get
\begin{equation}
\exists x \forall y (R_B (x) \land [\neg R_B (y) \lor (x = y)]
\end{equation}

Compiling this formula into tensor notation is rather straightforward. 
\begin{equation}
\resizebox{\linewidth}{!}{$
    \displaystyle
 \mathcal{T}_{\text{one-}B} =\\ min_1 \Big(\sum_{i=1}^{N} 1 - min_1\Big( \sum_{j=1}^{N} \mathcal{R}^b \mathbf{e}_i \bullet [(1- \mathcal{R}^b \mathbf{e}^j) +\\
 +(\mathbf{e}_i\bullet \mathbf{e}_j)]\Big)\Big)$}   
\end{equation}

Intuitively, this formula checks whether two domain elements are the same via the inner product, and if both domain elements have the property of being a $b$, then the formula evaluates to 1. If either formula is not a $b$, or if two different domain emelents are a $b$, the formula evaluates to 0. We can apply this to to the successor model for $abba$ in (\ref{eqn:succabba}) by defining each of the relational tensors in the formula over the domainand relations in the model. Doing so, it is easy to see that the formula evaluates to 0 (false) for domain elements 2 and 3, which are distinct and each have the property of being a $b$. 

\subsection{Compiling a Non-Counting Formula}
Next we demonstrate a formula which is properly First Order over the precedence model for strings, characterizing a Non-Counting language. We motivate this formula using a phonological pattern from Latin, in which in certain cases an $l$ cannot follow another $l$ unless an $r$ intervenes, no matter the distance between them \cite{Jensen1974,Heinz-2010-LLP}. This can be seen in the $-alis$ adjectival suffix  which appears as $-aris$ if the word
it attaches to already contains an $l$, except in cases where there is an intervening $r$, in which it appears again as -alis. 


The blocking effects in this non-local alternating pattern requires the use of quantifiers, and is properly Non-Counting. We can represent it with the following first-order formula:
\begin{multline}
F_{diss} = \forall x \forall y [R_l (x) \land R_l (y) \land R_\prec (x,y)] \rightarrow \\ \rightarrow \exists z [R_r (z) \land R_\prec (x,z) \land R_\prec (z,y)]
\end{multline}
Converting this into prenex normal form we get
\begin{multline}
    \exists x \exists y \exists z \neg [ R_l(x) \land R_l(y) \land R_\prec (x,y)] \lor \\
    \lor [R_r (z) \land R_\prec (x,z) \land R_\prec (z,y)]
\end{multline}

The compilation of this formula is again quite straightforward:

 \begin{multline}
\mathcal{T}_{diss} = min_1\Big( \sum_{i=1}^{N}min_1\Big( \sum_{j=1}^{N}min_1\Big( \sum_{k=1}^{N} \\
1 - \Big[(\mathcal{R}^l \mathbf{e}_i) \bullet  (\mathcal{R}^l \mathbf{e}_j)  \bullet (\mathbf{e}_i\mathcal{R}^\prec \mathbf{e}_j) \Big] +\\
+\Big[ (\mathcal{R}^z \mathbf{e}_k) \bullet (\mathbf{e}_i\mathcal{R}^\prec \mathbf{e}_k) \bullet (\mathbf{e}_k\mathcal{R}^\prec \mathbf{e}_j)\Big]\Big)\Big)\Big)
\end{multline}
Intuitively, this formula tests whether, for any two domain elements labeled $l$ and precede each other, there is another element labeled $r$ which comes between them. The use of the precedence relation here shows that this can happen anywhere in the word, and can thus handle the Latin dissimilation patterns above.

\section{Extension to Tree Structures}

The model-theory framework also allows describing tree structures. \cite{rogers2003syntactic} describes a model-theoretic characterization of trees of arbitrary dimensionality. In his framework, we specify the domain $D$ as a Gorn tree domain \cite{gorn1967explicit}.  This is a hereditarily prefix closed set \textit{D} of node addresses, that is to say, for every $d \in D$ with $d \coloneqq \alpha i$, it holds that $\alpha \in D$, and for every $d \in D$ with $d \coloneqq \alpha i \neq \alpha0,$ then $\alpha(i-1) \in D$. In this view, a string may be called a one-dimensional or unary-branching tree, since it has one axis along which its nodes are ordered. In a standard tree, on the other hand, the set of nodes is ordered as above by two relations, ``dominance" and ``immediate left-of". Suppose $s$ is the mother of two nodes $t$ and $u$ in some standard tree, and also assume that $t$ precedes $u$. Then we might say that $s$ dominates the string $tu$. 

\begin{figure}[htbp!]
\centering
\begin{tikzpicture}
	[level 1/.style={sibling distance=24mm,level distance=12mm},
	 level 2/.style={sibling distance=12mm}]	
	\node {$\epsilon$}
		child {node  (D){0} 
			child {node (E){00}}
			child {node (F){01}
				child {node (G){010}}
				child {node (H){011}}
				}
			}
		child {node (I){1}
			child {node (J){10}}
			child {node (M){11}
				child {node (N){110}}
				child {node (O){111}
					child {node (L){1110}}
					}
				child {node (P){112}}
				}
			};

\draw[dotted] (D) -- (I);
\draw[dotted] (E) -- (F);
\draw[dotted] (J) -- (M);
\draw[dotted] (G) -- (H);
\draw[dotted] (N) -- (O);
\draw[dotted] (O) -- (P);

\end{tikzpicture}

\caption{2-dimensional tree model. Dominance and precedence relations shown with solid/dashed and dotted lines, respectively}
\label{fig:3dtree}
\end{figure}
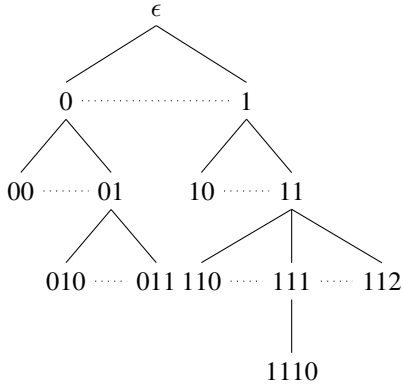

While a Gorn tree domain as written encodes these dominance and precedence relations implicitly, we may explicitly write them out model-theoretically so that a signature for a $\Sigma$-labeled 2-$d$ tree $T$ is $\Gamma^\succ=\tuple{D;\succ,\prec, [R_\sigma]_{\sigma\in\Sigma} }$ where $\succ$ is the immediate dominance relation and $\prec$ is the ``immediate left-of" relation. Model signatures that include transitive closure relations of each of these have also been studied.

We can thus generalize strings as 1-dimensional tree models, and standard trees as two-dimensional trees, which relate nodes to one-dimensional trees by immediate dominance. A three-dimensional tree relates nodes to two-dimensional, i.e. standard trees.  
In general, a $d$-dimensional tree is a set of nodes ordered by $d$ dominance relations such that the $n$-th dominance relation relates nodes to $(n-1)$-dimensional trees (for $d = 1$, single nodes are zero-dimensional trees). Importantly, we may compile these into tensors without the recursive role embeddings in Smolensky's formulation.

\section{Conclusion}
This paper provided a method for geometrically characterizing subregular languages in vector spaces. Model-theoretic descriptions of relational structures were embedded in Euclidean vector spaces, and statements in first-order logic over these structures were compiled into tensor formulas. Semantic evaluation was given via tensor contraction over tensors implementing a specific model. This method can easily be extended to consider other relational structures, and to other logics. Another application is to consider logical translations between model signatures, which define mappings between structures \cite{courcelle1994}, another area relevant for linguistics. The analytical power given by multilinear algebra, combined with the representational flexibility given by finite model theory and mathematical logic, provides a powerful combination for analyzing the nature of linguistic structures and cognition, and for exploring the relationship of languages and computation more generally.

\bibliographystyle{named}
\bibliography{ijcai}

\end{document}